\documentclass[]{frank}     

\usepackage{amsmath}
\usepackage{amscd, amsthm, amsfonts, amssymb}
\usepackage{mathrsfs, bbm, cancel, xfrac}
\usepackage{marvosym, enumerate, enumitem, upgreek}
\usepackage[all]{xy}
\usepackage[normalem]{ulem}
\usepackage{graphicx}
\usepackage{hyperref, color, microtype}
\usepackage{rotating, lscape, comment}
\definecolor{darkblue}{rgb}{0,0,1}
\hypersetup{pdfpagemode=UseNone, colorlinks=true, breaklinks=true, linkcolor=darkblue, menucolor=black, urlcolor=darkblue, citecolor=red}

\theoremstyle{remark}

\theoremstyle{definition}
\newtheorem{proposition}{Proposition}

\newtheorem{corollary}[proposition]{Corollary}
\newtheorem{remark}[proposition]{Remark}
\newtheorem{example}[proposition]{Example}

\newcommand{\RR}{\mathbbm R}

\newcommand{\CC}{\mathbbm C}
\newcommand{\HH}{\mathbbm H}

\newcommand{\KK}{\mathbbm{K}}

\newcommand{\cl}{{\rm C}\ell}

\def\DHLhksqrt#1#2{%
\setbox0=\hbox{$#1\sqrt{#2\,}$}\dimen0=\ht0
\advance\dimen0-0.2\ht0
\setbox2=\hbox{\vrule height\ht0 depth -\dimen0}%
{\box0\lower0.4pt\box2}}

\allowdisplaybreaks
\renewcommand{\arraystretch}{1.5}

\begin{document}

\sloppy \raggedbottom

\title{An explicit description of $SL(2,\CC)$ in terms of $SO^+(3,1)$ and vice versa
}
	\runningheads{$SL(2,\CC)$ in terms of $SO^+(3,1)$ and vice versa}
		{F.~Klinker}

\begin{start}

	\author{Frank Klinker}{}

	\address{Faculty of Mathematics, TU Dortmund University, 44221 Dortmund, Germany\\[0.5ex]
		\href{mailto:frank.klinker@math.tu-dortmund.de}{frank.klinker@math.tu-dortmund.de}	\\ }{}

\noindent
\makebox[0.8\textwidth]{%
\begin{minipage}{0.85\textwidth}
\begin{Abstract} 
In this note we present explicit and elementary formulas for the correspondence between the group of special Lorentz transformation $SO^+(3,1)$, on the one hand, and its spin group $SL(2,\CC)$, on the other hand.
\\
Although we will not mention Clifford algebra terminology explicitly, it is hidden in our calculations by using complex $2\times2$-matrices. Nevertheless, our calculations are strongly motivated by the Clifford algebra $\mathfrak{gl}(4,\CC)$ of four-dimensional space-time.
	\end{Abstract}
\end{minipage}}
\end{start}


\section{Introduction}

It is well known that for a pseudo-euclidean vector space $(V,g)$ the universal cover of the special orthogonal group $SO(V,g)$ is given by the so called spin group $Spin(V,g)$. For the case $V=\RR^{p+q}$ and $g={\rm diag}(\mathbbm{1}_q,-\mathbbm{1}_p)$ we write $SO(p,q)$ and $Spin(p,q)$. 
\renewcommand{\thefootnote}{}\footnotetext{{\em Int.~Electron.~J.~Geom.} {\bf 8} (2015) no.~1, 94-104}\renewcommand{\thefootnote}{\arabic{footnote}} The covering map is 2:1 for $\dim V>2$. The theoretic setting in which spin groups and related structures are best described is the Clifford algebra $\cl(V,g)$, see \cite{Che,Har,MichLaw} for example. Although spin groups in general refrain from being described by classical matrix groups for dimensional reason, there are accidental isomorphisms to such in dimension three to six, see Table \ref{tab:low}. The isomorphisms are a consequence of the classification of Lie algebras and can for example be seen by recalling the connection to Dynkin diagrams. We use the notation from \cite{Helga} and recommend this book for details on the definition of the classical matrix groups.
Due to the fact that the complexifications of the orthogonal groups are independent of the signature of the pseudo-Riemannian metric the groups in each column of Table \ref{tab:low} are real forms of the same complex group for fixed dimension.
\begin{table}[htb]\caption{The isomorphisms in dimensions $3\leq p+q\leq 6$}
$\displaystyle
{\renewcommand{\arraystretch}{1.5}
\begin{array}{c|c||c|c||c|c||c|c}
\multicolumn{2}{c||}{p+q=3} & \multicolumn{2}{c||}{p+q=4} 
		&\multicolumn{2}{c||}{p+q=5} & \multicolumn{2}{c}{p+q=6} \\\hline
(3,0)& SU(2)     &(4,0)& SU(2)^2     &(5,0)& Sp(2)     &(6,0)& SU(4)\\	       
(2,1)& SL(2,\RR) &(3,1)& SL(2,\CC)	 &(4,1)& Sp(1,1)   &(5,1)& SL(2,\HH)\\
	&		     &(2,2)& SL(2,\RR)^2 &(3,2)& Sp(4,\RR) &(4,2)& SU(2,2)\\
	&&&&&	     													 &(3,3)& SL(4,\RR)
\end{array}
}
$
\label{tab:low}
\end{table}

Infinitesimally, i.e.\ on Lie algebra level, the 2:1 covering structure cannot be seen. Therefore, the description on this level is given by  fixing  bases in the respective Lie algebras. If we try to take over this to the groups we see that the exponential map enters in the construction. A useful and manageable description is not obtained in general due to the Baker-Campbell-Hausdorff formula. However, in dimension four such description is possible and we present explicit formulas for the maps that connect $SO(3,1)$ and $SL(2,\CC)$. 

\section{Some preliminaries}

We will give some preliminaries on Lorentz transformations, Pauli matrices, $\mathfrak{gl}(2,\CC)$, and $SL(2,\CC)$, mainly to fix our notation.

By $\mathfrak{gl}_n\KK$ we denote the set of all ($n\times n$)-matrices over the field $\KK$ and by $GL(n,\KK)\subset\mathfrak{gl}_n\KK$ the group of all regular matrices.
The set of Lorentz transformations $O(3,1)$ by definition contains all elements $T\in GL(4,\RR)$ that obey $\|T\vec{x}\|^2=\|\vec{x}\|^2$ for $\vec{x}=(x^0,x^1,x^2,x^3)^t\in\RR^4$. Here we use 
\[
\|\vec{x}\|^2=(x^0)^2-(x^1)^2-(x^2)^2-(x^3)^2=\sum\limits_{i,j=0}^3g_{ij}x^ix^j\,,
\]
where 
\begin{equation}\label{minkmetric}
g=(g_{ij})_{i,j=0,\ldots,3}={\rm diag}(1,-1,-1,-1)
\end{equation} 
denotes the Minkowski metric and we write $\RR^{3,1}=(\RR^4,g)$. We denote the matrix entries of an endomorphism $T$ by $T^i{}_j$ such that $(T\vec{x})^i=\sum\limits_{j=0}^3T^i{}_jx^j$. 
\begin{remark}
The Lorentz transformations form a subgroup of $GL(4,\RR)$. As a submanifold of $GL(4,\RR)$ the group structure is smooth such that $SO(3,1)$ is indeed a Lie group. This follows also from a more general fact stating that closed subgroups of Lie groups are Lie subgroups, see \cite[Theorem II.2.3]{Helga}. 

$SO(3,1)$ admits four connected components that are associated to orientability and time-orientability of $\RR^{3,1}$. 
The connected component of the identity is given by the Lorentz transformations that obey ${\rm det}(T)=1$ and $T^0{}_0>0$. The special Lorentz transformations form a subgroup denoted by $SO^+(3,1)$.
\end{remark}
We use the following Pauli matrices:
\begin{equation}
\sigma_0 =  \begin{pmatrix}1&0\\0&1 \end{pmatrix},\quad
\sigma_1 =  \begin{pmatrix}0&1\\1&0 \end{pmatrix},\quad
\sigma_2 =  \begin{pmatrix}0&-i\\i&0 \end{pmatrix},\quad
\sigma_3 =  \begin{pmatrix}1&0\\0&-1 \end{pmatrix}\,.
\end{equation}
The matrices $\sigma_1,\sigma_2$, and $\sigma_3$ obey
\begin{equation}
\begin{aligned}
&\sigma_1^2=\sigma_2^2=\sigma^2_3=\sigma_0^2=\sigma_0\,, \\
&\sigma_1\sigma_2=-\sigma_2\sigma_1\,,\quad 
\sigma_1\sigma_3=-\sigma_3\sigma_1\,,\quad
\sigma_2\sigma_3=-\sigma_3\sigma_2\,,\\
&\sigma_1\sigma_2=i\sigma_3\,,\quad
\sigma_2\sigma_3=i\sigma_1\,,\quad
\sigma_3\sigma_1=i\sigma_2\,.
\end{aligned}\label{1}
\end{equation}
This can be combined to
\begin{equation}\label{c}
\sigma_i\sigma_j=\delta_{ij}\sigma_0+i\sum_{k=1}^3\epsilon_{ijk}\sigma_k
\end{equation}
with $\epsilon_{ijk}$ totally skew symmetric and $\epsilon_{123}=1$. In particular, 
for $i=0,1,2,3$ we have the nice relation 
\begin{equation}\label{orth}
\delta_{ij}=\frac{1}{2}{\rm tr}(\sigma_i\sigma_j)\,,
\end{equation}
and, moreover, from (\ref{c}) we get for $i=1,2,3$
\begin{equation}\label{d}
\sum_{j=1}^3 \sigma_j\sigma_i\sigma_j =-\sigma_i\ \text{ and }\ 
\sum_{j=0}^3 \sigma_j\sigma_i\sigma_j =0\,.
\end{equation}
We consider the natural $\RR$-linear map 
\begin{equation}\begin{aligned}
\Psi:\RR^4&\to\mathfrak{gl}(2,\CC)\,,\\
 \vec{x}=\begin{pmatrix}x^0\\x^1\\x^2\\x^3\end{pmatrix}&\mapsto \Psi(\vec{x})=\sum_{i=0}^3x^i\sigma_i = \begin{pmatrix}x^0+x^3 &x^1-ix^2 \\ x^1+ix^2 &x^0-x^3\end{pmatrix}\,.
\end{aligned}\label{psi}\end{equation}
The image $\Psi(\vec{x})$ of $\vec x\in\RR^4$ is a Hermitian matrix, i.e.
\[
\Psi(\vec{x})\in \mathfrak{h}(2,\CC):=\{A\in\mathfrak{gl}(2,\CC)\ |\ A=A^\dagger\}\,,
\] 
and, therefore, of type $\begin{pmatrix}a & \bar w\\\ w & b\end{pmatrix}$ with $a,b\in \RR$ and $w\in \CC$. The inverse map is given by
\[
\Psi^{-1}\left(\begin{pmatrix}a &\bar w\\ w & b\end{pmatrix}\right) =\begin{pmatrix}\frac{1}{2}(a+b) \\ Re(w) \\ Im(w)\\ \frac{1}{2}(a-b)\end{pmatrix}\,.
\]
In particular, each Hermitian matrix $B\in \mathfrak{h}(2,\CC)$ can be written as $B=x_0\sigma_0+x_1\sigma_1+x_2\sigma_2+x_3\sigma_3$ with $x_i\in\RR$ and we have 
\[
\|\vec{x}\|^2=(x^0)^2-(x^1)^2-(x^2)^2-(x^3)^2=\det(\Psi(\vec{x}))\,.
\]
As noticed in the title the special linear group $SL(2,\CC)\subset\mathfrak{gl}(2,\CC)$ will play an important role in the following and we will recall its definition:
\begin{equation}\label{sl}
SL(2,\CC)=\big\{ B\in\mathfrak{gl}(2,\CC) \,|\, \det(B)=1\big\}\,.
\end{equation}
$SL(2,\CC)$ is a Lie group, that has complex dimension three or real dimension six: in (\ref{sl}) we have one complex equation for the four complex parameters.
Each matrix in $SL(2,\CC)$ can be written in the form  
\begin{equation}\label{A}
A=a^0\sigma_0+a^1\sigma_1+a^2\sigma_2+a^3\sigma_3
\end{equation}
with
\begin{equation}\label{det}
\det(A)=(a^0)^2-(a^1)^2-(a^2)^2-(a^3)^2=1\,.
\end{equation}
Complex generators of $SL(2,\CC)$ are, for example, $\sigma_1,\sigma_2$ and $\sigma_3$.\footnote{Because of (\ref{1}) we have $\exp(b^1\sigma_1+b^2\sigma_2+b^3\sigma_3)=a^0+a^1\sigma_1+a^2\sigma_2+a^3\sigma_3$ with complex coefficients $a^i$ which depend on the $b^j$. This follows from a more general relation between Lie groups and their tangent space at the identity, i.e.~their Lie algebra, see \cite[Proposition II.1.6]{Helga}.}
Therefore, real generators are $\sigma_1$, $\sigma_2$, and $\sigma_3$, as well as $i\sigma_1$, $i\sigma_2$, and $i\sigma_3$.

If we omit in (\ref{A}) the condition on the determinant we get all of $\mathfrak{gl}(2,\CC)$ by such linear combination. The product of two matrices $A=\sum\limits_{j=0}^3a^j\sigma_j$ and $B=\sum\limits_{j=0}^3b^j\sigma_j$ expands as
\begin{equation}\label{p}
AB= \sum_{j=0}^3a_jb_j\sigma_0 +\sum_{j=1}^3(a_0b_j+b_0a_j)\sigma_j + i\sum_{j,k,\ell=1}^3\epsilon_{jk\ell}\,a_jb_k\,\sigma_\ell\,.
\end{equation}
Given a matrix $A\in\mathfrak{gl}(2,\CC)$ we define the conjugated matrix by 
\begin{equation}\label{bar}
 A' =  a^0\sigma_0- a^1\sigma_1- a^2\sigma_2- a^3\sigma_3 \,.
\end{equation}
This conjugate obeys ${\rm det}(A')={\rm det}(A)$ and $A'B'=(BA)'$. In particular, the product of a matrix and its conjugated is given by 
\begin{equation}
A' A= A A' = \big((a^0)^2-(a^1)^2-(a^2)^2-(a^3)^2\big)\sigma_0\,,
\end{equation}
such that its trace obeys 
\begin{equation}\label{Atrace}
\frac{1}{2}{\rm tr}(A'A)= \det(A)=(a^0)^2-(a^1)^2-(a^2)^2-(a^3)^2\,.
\end{equation}
Moreover, for $A\in SL(2,\CC)$ we have $A'\in SL(2,\CC)$ and 
$A^{-1}= A'$ due to (\ref{det}).

We collect the symmetry properties (\ref{1}) and the symmetry property (\ref{bar}) as follows. We introduce signs $\varepsilon_i$ and $\varepsilon_{ij}$ defined by $\sigma_i'=\varepsilon_i\sigma_i$ and $\sigma_i\sigma_j=\varepsilon_{ij}\sigma_j\sigma_i$, i.e.
\begin{equation}\label{signs}
\big(\varepsilon_i\big)_{i=0,\ldots,3}=\begin{pmatrix}
1\\-1\\-1\\-1
\end{pmatrix}\,,\quad
\big(\varepsilon_{ij}\big)_{i,j=0,\ldots,3}=\begin{pmatrix}
1 & 1 & 1 & 1\\
1 & 1 & -1 & -1\\
1 & -1 & 1 & -1\\
1 & -1 & -1 & 1
\end{pmatrix}\,.
\end{equation}
In particular, in terms of $\varepsilon_i$ Minkowski metric (\ref{minkmetric}) reads as
\[
g_{ij}=\varepsilon_i\delta_{ij}=\varepsilon_j\delta_{ij}\,.
\]

\section{$SO(3,1)$ in terms of $SL(2,\CC)$}

We consider an action $\Phi$ of $SL(2,\CC)$ on the set of Hermitian matrices that is defined by
\begin{equation}
SL(2,\CC)\times \mathfrak{h}(2,\CC) \ni (A,B)\mapsto ABA^\dagger \in \mathfrak{h}(2,\CC)\,.
\end{equation}
Writing  $B=\Psi(\vec{x})$ the combination $\Psi^{-1}( A\Psi(\vec{x}) A^\dagger )$
yields an element in $\RR^{4}$. This defines an action $\Phi$ of $SL(2,\CC)$ on $\RR^{4}$ via
\begin{equation}\label{X}
\Phi(A) (\vec{x})= \Psi^{-1}(A\Psi(\vec{x})A^\dagger)
\end{equation}
The map $\Phi(A):\RR^{4}\to\RR^{4}$ is $\RR$-linear.
Furthermore, we have
\begin{align*}
\|\Phi(A)(\vec{x})\|^2& = \|\Psi^{-1}( A\Psi(\vec{x}) A^\dagger )\|^2
= \det (A\Psi(\vec{x}) A^\dagger) \\
& =\det(A)\, \overline{\det(A)}\, \det(\Psi(\vec{x})) = \det (\Psi(\vec{x}))=\|\vec{x}\|^2
\end{align*}
such that $\Phi(A)$ is a Lorentz transformation.

The Matrix entries of $T:=(\Phi(A)^i{}_j)_{i,j=0,\ldots,3}$ depend on the complex parameters $a_i$ from the decomposition of $A$ according to (\ref{A}). They can explicitly be expressed by expanding and rearranging the right hand side of (\ref{X}):
\begin{align}
\Psi\big(\Phi(A)(\vec{x})\big)=\ &\sum_{i,j=0}^4T^i{}_jx^j\sigma_i \nonumber\\
=\ &\quad \, 
	 \big( a^0\bar a^0 + a^1\bar a^1 + a^2\bar a^2  + a^3\bar a^3\big)  x^0\sigma_0 \nonumber\\
   &+\big( a^0\bar a^1 + a^1\bar a^0 - ia^2\bar a^3 + ia^3\bar a^2\big) x^1\sigma_0 \nonumber\\
   &+\big( a^0\bar a^2 + a^2\bar a^0 + ia^1\bar a^3 - ia^3\bar a^1\big) x^2\sigma_0 \nonumber\\
   &+\big( a^0\bar a^3 + a^3\bar a^0 - ia^1\bar a^2 + ia^2\bar a^1\big) x^3\sigma_0 \nonumber\\ 
   &+\big( a^0\bar a^1 + a^1\bar a^0 + ia^2\bar a^3 - ia^3\bar a^2\big) x^0\sigma_1 \nonumber\\
   &+\big( a^0\bar a^0 + a^1\bar a^1 - a^2\bar a^2  - a^3\bar a^3\big)  x^1\sigma_1 \nonumber\\
   &+\big( a^1\bar a^2 + a^2\bar a^1 + ia^0\bar a^3 - ia^3\bar a^0\big) x^2\sigma_1 \nonumber\\
   &+\big( a^1\bar a^3 + a^3\bar a^1 - ia^0\bar a^2 + ia^2\bar a^0\big) x^3\sigma_1 \label{schwer}\\
   &+\big( a^0\bar a^2 + a^2\bar a^0 - ia^1\bar a^3 + ia^3\bar a^1\big) x^0\sigma_2 \nonumber\\
   &+\big( a^1\bar a^2 + a^2\bar a^1 - ia^0\bar a^3 + ia^3\bar a^0\big) x^1\sigma_2 \nonumber\\
   &+\big( a^0\bar a^0 - a^1\bar a^1 + a^2\bar a^2  - a^3\bar a^3\big)  x^2\sigma_2 \nonumber\\
   &+\big( a^2\bar a^3 + a^3\bar a^2 + ia^0\bar a^1 - ia^1\bar a^0\big) x^3\sigma_2 \nonumber\\
   &+\big( a^0\bar a^3 + a^3\bar a^0 + ia^1\bar a^2 - ia^2\bar a^1\big) x^0\sigma_3 \nonumber\\
   &+\big( a^1\bar a^3 + a^3\bar a^1 + ia^0\bar a^2 - ia^2\bar a^0\big) x^1\sigma_3 \nonumber\\
   &+\big( a^2\bar a^3 + a^3\bar a^2 - ia^0\bar a^1 + ia^1\bar a^0\big) x^2\sigma_3 \nonumber\\
   &+\big( a^0\bar a^0 - a^1\bar a^1 - a^2\bar a^2  + a^3\bar a^3\big)  x^3\sigma_3 \nonumber\,.
\end{align}
By applying (\ref{orth}) directly to (\ref{X}) we see that (\ref{schwer}) gets the following compact form.
\begin{proposition}\label{hat}
Consider $A\in SL(2,\CC)$ and $\Phi:SL(2,\CC)\to SO(3,1)$. Then the image of $T=\Phi(A)$ has the entries
\begin{equation}\label{leicht}
\fbox{$\displaystyle T^i{}_j = \frac{1}{2}{\rm tr}(\sigma_i A\sigma_j A^\dagger)= \frac{1}{2}{\rm tr}( A\sigma_j A^\dagger\sigma_i)$}
\end{equation}
\end{proposition}
\begin{remark}
As we saw, the map $\Phi(A)$ is a Lorentz transformation -- but is it a special Lorentz transformation as well? From (\ref{schwer}) we see directly that $\Phi(A)^0{}_0>0$ -- but what about the determinant of $\Phi(A)$? Without calculating the determinant we can see the result as follows: 
The image of the map $\Phi: SL(2,\CC)\to O(3,1)$ is connected because $\Phi$ is continuous and $SL(2,\CC)$ is (simply) connected. Furthermore, the identity is in the image of $\Phi$ such that all of the image of $\Phi$ is contained in $SO^+(3,1)$.
\end{remark}

\section{$SL(2,\CC)$ in terms of $SO(3,1)$}

By explicitly inverting the system (\ref{schwer})we show  in this section that for any special Lorentz transformation  $T\in SO^+(3,1)$ their exists a matrix $A\in SL(2,\CC)$ with $T=\Phi(A)$. 
This matrix isn't unique, because with $A$ its negative $-A\in SL(2,\CC)$ obeys $\Phi(-A)=\Phi(A)$, too. 
This 2:1 behavior will be reflected in the existence of a square root during the process of solving the equation  $T=\Phi(A)$, i.e.~(\ref{schwer}), for $a^i$.

To solve (\ref{schwer}) we write $\Phi(A)=T$ and consider (\ref{X}) and (\ref{leicht}),  i.e.~$\Psi(T\vec{x})= A\Psi(\vec{x}) A^\dagger $ and $T^i{}_j=\frac{1}{2}{\rm tr}(\sigma^iA\sigma_jA^\dagger)$. We define matrices $\tau_{(i)}(T)\in \mathfrak{gl}(2,\CC)$ by
\begin{equation}\label{t}
\tau_{(i)}(T):=\sum_{j,k=0}^3T^j{}_k\sigma_i'\sigma_j\sigma_i\sigma_k\,.
\end{equation}
For example, $i=0$ yields
\begin{equation}
\tau_{(0)}(T)
   = \sum_{i=0}^3 T^i{}_i\sigma_0+\sum_{i=1}^3(T^i{}_0 +T^0{}_i)\sigma_i+i\sum_{i,j,k=1}^3T^i{}_j\epsilon_{ijk}\sigma_k \,.
\end{equation}
Moreover, due to (\ref{X}), we have
\begin{equation}\begin{aligned}
\tau_{(i)}(T) =\ & \sum_{k=0}^3\sigma_i'\big(\sum_{j=0}^3T^j{}_k\sigma_j\big)\sigma_i\sigma_k 
 = \sum_{k=0}^3 \sigma_i'\Psi( T(\vec{e}_k) )\sigma_i\sigma_k \\
 =\ & \sum_{k=0}^3 \sigma_i' A\sigma_k A^\dagger\sigma_i\sigma_k\,.
\end{aligned}\end{equation}
For arbitrary $B=\sum\limits_{i=0}^3b^i\sigma_i$ we get 
\begin{equation}
\sum_{j=0}^3 \sigma_j B\sigma_j 
=b^0\sum_{j=0}^3(\sigma_j)^2\sigma_0 +\sum_{i=1}^3b^i\sum_{j=0}^3\sigma_j\sigma_i\sigma_j
=4b^0 \sigma_0
\end{equation}
by using (\ref{d}). Doing the same calculations for $B\sigma_1$, $B\sigma_2$, and $B\sigma_3$ we get  for $i=0,1,2,3$
\begin{equation}
\sum_{j=0}^3 \sigma_j B\sigma_i\sigma_j  =  4b^i \sigma_0 
\end{equation}
such that 
\begin{equation}
\tau_{(i)}(T) = \sigma_i'A \sum_{j=0}^3\sigma_j A^\dagger\sigma_i\sigma_j =4\bar a^i \sigma_i'A\,.
\end{equation}
We use (\ref{bar}) and consider the following trace 
\begin{equation}
\begin{aligned}
\frac{1}{2}{\rm tr}\big(\tau_{(i)}(T)' \tau_{(i)}(T)\big) 
& = \frac{1}{2} {\rm tr} \big((4\bar a^i\sigma_i'A)'(4\bar a^i \sigma_iA) \big)\\
& = 8(\bar a^i)^2{\rm tr}(A'\sigma_i\sigma_i A)\\
& = 16(\bar a^i)^2 \,.
\end{aligned}
\end{equation}
This special combination of $\tau_{(i)}(T)$ and $\tau_{(i)}(T)'$ yields the following statement.
\begin{proposition}\label{check}
Consider $T\in SO^+(3,1)$. Then there exist maps \[\hat\Phi^\pm_{(i)}: SO^+(3,1)\to SL(2,\CC)\] for $i=0,1,2,3$ such that the images $\hat\Phi^\pm_{(i)}(T)$ are the solutions of $\Phi(A)=T$ if ${\rm tr}\big(\tau_{(i)}(T)'\tau_{(i)}(T)\big)\neq 0$. The maps are given by 
\begin{equation}
\fbox{$\displaystyle
\hat\Phi^\pm_{(i)}(T)=
 \pm\frac{1}{\sqrt{\frac{1}{2} {\rm tr}\big(\tau_{(i)}(T)'\tau_{(i)}(T)\big)}}\, \sigma_i'\tau_{(i)}(T)
$}  \label{sol}
\end{equation}
\end{proposition}

We will formulate the result (\ref{sol}) in terms of $T$ alone, i.e.\ without help of the map $\tau_{(i)}$. By using the signs (\ref{signs}) a more explicit way to express (\ref{t}) is  
\begin{align*}
\varepsilon_i\tau_{(i)}(T) & = \sum_{j,k=0}^3 T^j{}_k\sigma_i\sigma_j\sigma_i\sigma_k\\
		&=  \Big(\sum_{j=0}^3\varepsilon_{ij}T^j{}_j\Big)\sigma_0+\sum_{j=1}^3\big(\varepsilon_{ij}T^j{}_0+T^0{}_j\big)\sigma_j 
			+i\sum_{j,k,\ell=1}^3\varepsilon_{ij}\epsilon_{jk\ell}T^j{}_k\sigma_\ell\,.
\end{align*}
We write $T_{(i)}$ for the matrix with entries $(T_{(i)}){}^j{}_k=\epsilon_{ij}T^j{}_k$, in particular $T_{(0)}=T$.

For the expansion  $\varepsilon_i\tau_{(i)}=t^0\sigma_0+t^1\sigma_1+t^2\sigma_2+t^3\sigma_3$ we have
\begin{align*}
\frac{1}{2}{\rm tr}\big(\tau_{(i)}(T)'\tau_{(i)}(T)\big) 
=\ &
(t^0)^2-(t^1)^2-(t^2)^2-(t^3)^2 \\
=\ & \big({\rm tr}(T_{(i)})\big)^2-\sum_{j=1}^3\Big( 
		\varepsilon_{ij} T^j{}_0 +T^0{}_j
		+i\sum_{k,\ell=1}^3\varepsilon_{ik}\epsilon_{k\ell j}T^k{}_\ell\Big)^2\\
=\ & \big({\rm tr}(T_{(i)})\big)^2 
	 -\sum_{j=1}^3\big( T^j{}_0T^j{}_0 + T^0{}_jT^0{}_j 
	 + 2\varepsilon_{ij} T^j{}_0T^0{}_j\big) \\	
 & 	 -\!\!\!\sum_{j,k,\ell,m,n=1}^3\!\!\!\varepsilon_{ik}\varepsilon_{im}\epsilon_{k\ell j}
 			\epsilon_{mnj}T^k{}_\ell T^m{}_n  \\
 &	 +2i \sum_{j,k,\ell=1}^3\epsilon_{k\ell j}(\varepsilon_{ij}T^j{}_0+T^0{}_j)\varepsilon_{ik}
 							T^k{}_\ell\\
=\ & \big({\rm tr}(T_{(i)})\big)^2 -\sum_{i=1}^3\big( 
	 T^j{}_0T^j{}_0 + T^0{}_jT^0{}_j  + 2\varepsilon_{ij}T^j{}_0T^0{}_j\big) \\
 &   + \sum_{j,k=1}^3 T^j{}_kT^j{}_k 
 	 -\sum_{j,k=1}^3  \varepsilon_{ik}\varepsilon_{ij}T^j{}_kT^k{}_j \\
 &	 -2i\!\! \sum_{j,k,\ell=1}^3\!\!\epsilon_{jk\ell}(\varepsilon_{ij}T^j{}_0+T^0{}_j)\varepsilon_{ik}T^k{}_\ell\,,
\end{align*}
where we used $\sum_{\ell=1}^3\epsilon_{jk\ell}\epsilon_{mn\ell}=2\delta^{jk}_{mn}$ in the last step.
From the fact that $T$ is a Lorentz transformation we have 
\begin{equation}\label{covariant}
\sum_{i,k=0}^3T^i{}_jg_{ik}T^k{}_\ell=g_{j\ell}\,.
\end{equation}
After considering $j=\ell$ and multiplying by the sign $\varepsilon_j$ we take the sum over $j$ and obtain 
\begin{equation}\label{trace}
\sum_{k,j=1}^3T^k{}_jT^k{}_j-\sum_{j=1}^3T^j{}_0T^j{}_0-\sum_{j=1}^3T^0{}_jT^0{}_j=4-(T^0{}_0)^2\,.
\end{equation}
We use this and insert the positive signs $\varepsilon_{i0}$ to simplify
\begin{align*}
\frac{1}{2}{\rm tr}(\tau_{(i)}'\tau_{(i)})
=\ & 4 + \big({\rm tr}(T_{(i)})\big)^2
	 -2i\!\! \sum_{j,k,\ell=1}^3\!\!\epsilon_{jk\ell}(\varepsilon_{ij}T^j{}_0+T^0{}_j)\varepsilon_{ik}T^k{}_\ell\\
 &	 -\epsilon_{i0}\epsilon_{i0}T^0{}_0T^0{}_0 - 2\sum_{i=1}^3\varepsilon_{i0}\varepsilon_{ij}T^j{}_0T^0{}_j 
	 -\sum_{j,k=1}^3  \varepsilon_{ik}\varepsilon_{ij}T^j{}_kT^k{}_j \\
=\ & 4 + ({\rm tr}\big(T_{(i)})\big)^2
     -\sum_{j,k=0}^3  \varepsilon_{ik}\varepsilon_{ij}T^j{}_kT^k{}_j \\
& 	 -2i\!\! \sum_{j,k,\ell=1}^3\!\!\epsilon_{jk\ell}(\varepsilon_{ij}T^j{}_0+T^0{}_j)\varepsilon_{ik}T^k{}_\ell\\
=\ & 4 + \big({\rm tr}(T_{(i)})\big)^2 - {\rm tr}(T_{(i)}^2) 
 	 -2i\!\! \sum_{j,k,\ell=1}^3\!\!\epsilon_{jk\ell}(\varepsilon_{ij}T^j{}_0+T^0{}_j)\varepsilon_{ik}T^k{}_\ell\,.
\end{align*}
This yields the following Corollary of Proposition \ref{check} that contains the announced explicit description of $SL(2,\CC)$ in terms of $SO(3,1)$.
\begin{corollary}
In terms of the entries of $T$ formula (\ref{sol}) reads
\begin{equation}\tag{\ref{sol}${}_0$}
\hat\Phi^\pm_{(0)}(T) =\pm \frac{\displaystyle
	 	 {\rm tr}(T)\sigma_0
		+\sum\limits_{j=1}^3\left( T^j{}_0 +T^0{}_j 
		+i\sum\limits_{k,\ell=1}^3T^k{}_\ell\epsilon_{jk\ell}\right)\sigma_j
		}{
		\sqrt{4+ ({\rm tr}(T))^2 - {\rm tr}(T^2) 
		- 2i \sum\limits_{j,k,\ell=1}^3\epsilon_{jk\ell}(T^j{}_0+T^0{}_j)T^k{}_\ell}
		}\,
\end{equation}
for $i=0$, as well as 
\begin{equation}\tag{\ref{sol}${}_i$}
\begin{aligned}
\hat\Phi^\pm_{(i)}(T)
=  & \pm \frac{1}{
		\sqrt{4+ \big({\rm tr}(T_{(i)})\big)^2 - {\rm tr}(T_{(i)}^2) 
		- 2i \sum\limits_{j,k,\ell=1}^3\epsilon_{jk\ell}
			(\varepsilon_{ij}T^j{}_0+T^0{}_j)\varepsilon_{ik}T^k{}_\ell}
		}\times\\
 	&\qquad \times\left( 
 		\big( T^i{}_0 +T^0{}_i +i\!\!\sum\limits_{j,k=1}^3\!\!	
			\varepsilon_{ij}\epsilon_{ijk}T^j{}_k\big) \sigma_0
		+{\rm tr}(T_{(i)})\sigma_i \right.\\
	&\qquad \qquad \left.	+\sum\limits_{j=1}^3\Big(
		T^i{}_j-\varepsilon_{ij}T^j{}_i 
		+i\sum\limits_{k=1}^3\epsilon_{ikj}(\varepsilon_{ik} T^k{}_0 +T^0{}_k )
  		\Big)\sigma_j\right)			
\end{aligned}	
\end{equation}
for $i=1,2,3$.
\end{corollary}
\begin{remark}
All four combinations in (\ref{t}) are needed to describe full $SL(2,\CC)$ because formula (\ref{sol}) only works for $\tau_{(i)}(T)\neq0$. For example, the choice $\hat\Phi_0$ only works for matrices $T$ such that $a^0\neq 0$. 
In particular, the matrices $\sigma_1$, $\sigma_2$, and $\sigma_3$ that correspond to $T={\rm diag}(1,1,-1,-1)$,  $T={\rm diag}(1,-1,1,-1)$, and $T={\rm diag}(1,-1,-1,1)$, respectively, cannot be described.
\end{remark}
\begin{example}
We consider $T\in SO^+(3,1)$ such that
\[\begin{gathered}
T_{(0)}=T={\begin{pmatrix}
\cosh\alpha &\sinh\alpha & \\ \sinh\alpha &\cosh\alpha &\\&&\mathbbm{1}_2 
\end{pmatrix}}\,,\quad
T_{(1)}={\begin{pmatrix}
\cosh\alpha &\sinh\alpha & \\ \sinh\alpha &\cosh\alpha &\\&&-\mathbbm{1}_2
\end{pmatrix}}\,,\\
T_{(2)}={\begin{pmatrix}
\cosh\alpha &-\sinh\alpha  &\\ \sinh\alpha &-\cosh\alpha &\\&&\sigma_3
\end{pmatrix}}\,,\quad
T_{(3)}={\begin{pmatrix}
\cosh\alpha &-\sinh\alpha &\\ \sinh\alpha &-\cosh\alpha &\\&&-\sigma_3
\end{pmatrix}}\,.
\end{gathered}\]
Then
\[\begin{gathered}
{\rm tr}(T)=2(\cosh(\alpha)+1)\,,\quad
{\rm tr}(T_{(1)})=2(\cosh(\alpha)-1)\,, \\
{\rm tr}(T_{(2)})={\rm tr}(T_{(3)})=0\,. 
\end{gathered}\]
Furthermore we have 
\[
T^2=T_{(1)}^2=\begin{pmatrix}
2\cosh^2\alpha-1 &2\cosh\alpha\sinh\alpha & &\\2\cosh\alpha\sinh\alpha &2\cosh^2-1 & & \\& &\mathbbm{1}_2
\end{pmatrix},\quad T_{(2)}^2= T_{(3)}^2=\mathbbm{1}_4\,,
\]
and 
\begin{align*}
{\rm tr}(\tau_{(0)}(T)'\tau_{(0)}(T))&=8(\cosh2\alpha+1)\,,\\
{\rm tr}(\tau_{(1)}(T)'\tau_{(1)}(T))&=8(\cosh2\alpha-1)\,,\\
{\rm tr}(\tau_{(2)}(T)'\tau_{(2)}(T))& ={\rm tr}(\tau_{(3)}(T)'\tau_{(3)}(T))=0\,.
\end{align*}
Therefore, we can consider $\hat\Phi^\pm_{(0)}(T)$ and $\hat \Phi^\pm_{(1)}(T)$ and get
\begin{align*}
\hat\Phi^\pm_{(0)}(T)&=\frac{1}{\sqrt{2}}
	\left(\sqrt{\cosh\alpha+1}\,\sigma_0+\frac{\sinh\alpha}{\sqrt{\cosh\alpha+1}}\,\sigma_1\right)\,,\\
\hat\Phi_{(1)}^\pm(T)&=\frac{1}{\sqrt{2}}
	\left(\frac{\sinh\alpha}{\sqrt{\cosh\alpha-1}}\,\sigma_0+\sqrt{\cosh\alpha-1}\,\sigma_1\right).
\end{align*}
Both yield the same Matrix $A$ , namely
\[
A=\begin{pmatrix}
\cosh\frac{\alpha}{2}&\sinh\frac{\alpha}{2} \\
\sinh\frac{\alpha}{2}& \cosh\frac{\alpha}{2}
\end{pmatrix},
\]
which follows from
$2\sinh^2\frac{\alpha}{2}=\cosh\alpha-1$ and $2\cosh^2\frac{\alpha}{2}=\cosh\alpha+1$.
\end{example}
\section{Some concluding remarks}

\begin{itemize} 
\item
The results that we presented here in an elementary way have been discussed in parts in the literature. 
The particular choice $i=0$ in (\ref{sol}) has been discussed in \cite[p.~69]{Joo} where the author states a variant of formula (\ref{sol}${}_0$), in \cite[p.~53]{Hest} where the author emphasizes that the formula only holds in special cases, and in \cite[p.~130]{Lou} with reference to \cite{Hest} but without comment on the incompleteness.
\item
On purpose we neglected the use of the theory of Clifford algebras and their representations although there is a strong relation. 
In fact, the Clifford algebra $\cl(4)$ is the framework in which the results above can be formulated and we will shortly recall how Pauli matrices enter into the discussion.
Starting in dimension two we see that the set $\{\sigma_1,\sigma_2\}$ provides generators of the Clifford algebra $\cl(2)$ because $\sigma_i\sigma_j+\sigma_j\sigma_i=2\delta_{ij}$ for $i=1,2$.
By adding the volume element $\sigma_3=-i\sigma_1\sigma_2$ we get generators $\{\sigma_1,\sigma_2,\sigma_3\}$ of $\cl(3)$ because the same relations as before hold but for $1\leq i\leq 3$. 
As we can check the set $\{\Sigma_0=\sigma_1\otimes\mathbbm{1},\Sigma_1=\sigma_2\otimes\sigma_1,\Sigma_2=\sigma_2\otimes\sigma_2,\Sigma_3=\sigma_2\otimes\sigma_3,\}$ obeys $\Sigma_i\Sigma_j+\Sigma_j\Sigma_i=2\delta_{ij}$ for $0\leq i\leq 3$ such that it yields generators of $\cl(4)$. 
Such doubling process can always be used when going from $\cl(2k-2)$ to $\cl(2k)$. This gives a iterative way to construct $\cl(2k)$ from $\cl(2)$, see for example \cite{BFGK}. 
The doubling process, of course, is not unique because for any set of generators $\{\Sigma_i\}$ and any unitary transformation $\Omega$ the set $\{\Omega\Sigma_i\Omega^\dagger\}$ yields generators, too. 
Although we restricted to the complex Clifford algebra above, generators of the real Clifford algebra according to a metric with signature can easily be obtained by adding some extra $i$ in front of some of the generators. For more details we again refer to the literature, for example \cite{Che,Har,MichLaw}.
\item
Our choice for $\cl(4)$ above is the so called Weyl representation for which the subspace $\text{span}\big\{\Sigma_{ij}=\frac{1}{2}(\Sigma_i\Sigma_j-\Sigma_j\Sigma_i)\big\}\subset\cl(4)$ is block-diagonal. This subspace is isomorphic to the algebra of  skew-symmetric ($4\times4$)-matrices and reflects the algebra isomorphism $\mathfrak{so}(4)\simeq\mathfrak{so}(2)\oplus\mathfrak{so}(2)$. In terms of Dynkin diagrams this is $D_2=A_1\oplus A_1$ and here $\mathfrak{sl}(2,\CC)$ enters as the standard realization of $A_1$. A more geometric way to interpret the isomorphism is the notion of selfduality of two-forms in dimension four. In this particular dimension the Hodge operator provides an involution on the six-dimensional space of two-forms and, therefore, it splits into two three-dimensional eigenspaces, the so called self-dual and anti-self-dual two-forms.
\item
The introduction of the sign $\epsilon_i$ into (\ref{covariant}) to get (\ref{trace}) is somewhat artificial. In a more geometric way this is due to the natural isomorphism $\RR^4\simeq(\RR^4)^*$ defined by the Minkowski metric $g$. In terms of index-notation this is raising and lowering of indices.  
This isomorphism is needed when we want to calculate invariant traces of bilinear forms. In  fact, (\ref{covariant}) is a bilinear form rather than an endomorphism.
\item
There is a last nice relation we like to mention. The isomorphism $\Psi$ from (\ref{psi}) translates (\ref{p}) to $\RR^4$. After writing $\RR^4=\RR\times\RR^3$ this reflects the geometry of $\RR^3$, i.e.\ the Euclidean product $\langle\cdot,\cdot\rangle$ and the cross product $\times$. For this we write $\vec x=(x_0,\mathbf{x})$ with $\mathbf{x}\in\RR^3$. 
Then 
\[
\Psi^{-1}\big(\Psi(\vec x)\Psi(\vec y)\big)
=\begin{pmatrix}
x_0y_0+\langle\mathbf{x},\mathbf{y}\rangle\\ 
x_0\mathbf{y}+y_0\mathbf{x}+\mathbf{x}\times\mathbf{y}
\end{pmatrix},
\]
and therefore
\[
\tfrac{1}{2}\Psi^{-1}\big(\Psi(\vec x)\Psi(\vec y)-\Psi(\vec y)\Psi(\vec x)\big)
=\begin{pmatrix}
0\\ 
\mathbf{x}\times\mathbf{y}
\end{pmatrix}.
\]
\end{itemize}

\end{document}